\documentclass{ws-procs975x65}

\begin{document}
\title{COSMOLOGICAL k-ESSENCE CONDENSATION}

\author{NEVEN BILI\'C}

\address{Rudjer Bo\v{s}kovi\'{c} Institute, 10002 Zagreb, Croatia \\
E-mail:bilic@thphys.irb.hr}

\author{GARY B.~TUPPER$^*$ and RAOUL D.~VIOLLIER$^\dag$}

\address{Centre of Theoretical Physics and Astrophysics,\\
University of Cape Town,  Rondebosch 7701, South Africa\\
$^*$E-mail:gary.tupper@uct.ac.za; $^\ddag$E-mail:raoul.viollier@uct.ac.za
}


\begin{abstract}
We consider a model of dark energy/matter unification based 
on a k-essence type of theory similar to tachyon condensate models.
Using an extension of the general relativistic spherical model 
which incorporates the effects of both pressure and the acoustic horizon 
we show that an initially perturbative k-essence fluid evolves into a mixed system
containing cold dark matter like gravitational condensate in significant quantities.
\end{abstract}


\bodymatter


\section*{~}

The most popular cosmological models such as
 $\Lambda$CDM model and 
a quintessence-CDM model 
assume that DM and DE are distinct entities.
Another interpretation of the observational data is that
DM/DE are different
manifestations of a common structure. 
The first definite model of this type was proposed a few years ago 
\cite{kam9,bil7,bil8},
based upon the Chaplygin gas,
a perfect fluid obeying the equation of state
\begin{equation}
p =  - \frac{A}{\rho} \; ,
\label{eq001}
\end{equation}
which has been extensively studied for its
mathematical properties \cite{jack10}.
The general class of models, in which a unification of DM and DE
is achieved through a single entity, is often referred to as
{\em quartessence} \cite{mak11,mak12}.
Among other scenarios of unification that have recently been suggested,
interesting attempts are based on
the so-called {\em k-essence} \cite{chi13,sch14}, a scalar field with
noncanonical kinetic terms which was first introduced as a model for
inflation \cite{arm15}.

All models that unify DM and DE face the problem of
nonvanishing sound speed and the well-known Jeans instability.
Soon after the appearance of
\cite{kam9}
and \cite{bil7},
it was pointed out that the perturbative Chaplygin gas 
(for early work see \cite{fab19},
and more recently \cite{gor20}) 
is incompatible with the observed mass power spectrum \cite{sand21}
and microwave background \cite{cart22}.
Essentially, these results are a consequence of
a nonvanishing comoving acoustic horizon
\begin{equation}
d_s = \int  dt \, \frac{c_s}{a} \; \; .
\label{eq005}
\end{equation}
 The perturbations whose comoving size $R$ is larger than $d_s$ grow as
$\delta = (\rho - \bar{\rho})/\bar{\rho} \sim a$.
 As soon as $R<d_s$, the perturbations undergo damped oscillations.
 For the Chaplygin gas we have
$d_s \sim a^{7/2}/H_0$, where $H_0$ is the present day value of the Hubble parameter, reaching Mpc scales already at redshifts of order 10. 
However, 
 as soon as $\delta \simeq 1$ the linear perturbation theory cannot be trusted. 
A significant fraction of initial density perturbations
collapses in gravitationally bound structure - the condensate and
 the system evolves into 
a two-phase structure - a mixture of
CDM in the form of condensate and DE in the form of uncondensed gas.

The simple Chaplygin gas does not exhaust all the possibilities for 
quart\-essence. A particular case of k-essence \cite{arm15} is
 the string-theory inspired
 tachyon Lagrangian \cite{sen40}
\begin{equation}
{\cal L}  =
 - V(\varphi) \; \sqrt{1 - g^{\mu \nu}
 \, \varphi_{,\mu} \, \varphi_{,\nu} } \; ,
\label{eq01}
\end{equation}
where
\begin{equation}
X \equiv g^{\mu \nu} \varphi_{, \mu} \varphi_{, \nu} \, .
\label{eq1203}
\end{equation}
It may be shown that every tachyon condensate model
can be interpreted as a 3+1 brane moving 
in a 4+1 bulk \cite{bil39,bilic}.
Eq.\ (\ref{eq001})
is obtained 
using the stress-energy tensor $T_{\mu \nu}$ derived from the Lagrangian 
(\ref{eq01})
with $V(\varphi)$
replaced by a constant $\sqrt{A}$.

In a recent paper \cite{bilic} we have developed a fully relativistic version of the spherical model for studying the evolution of density perturbations even into the fully nonlinear regime. The formalism  is similar in spirit to \cite{bil34} and applicable to any k-essence model.  The key element is an approximate method for treating the effects of pressure gradients. Here we give a brief description of our method and its application
 to a unifying model based on the Lagrangian (\ref{eq01}) with a potential of the form
\begin{equation}
V(\varphi)=V_n \varphi^{2n} \; \; ,
\label{eq08}
\end{equation}
where $n$ is a positive integer. In the regime where structure formation takes place, this model effectively
behaves as the variable Chaplygin   gas \cite{guo43} with the equation of state (\ref{eq001}) in which
$A \sim a^{6n}$. As a result, the much smaller acoustic horizon $d_{s} \sim a^{(7/2+3n)}/H_{0}$ enhances condensate formation by two orders of magnitude over the simple Chaplygin gas. Hence this type of model
may salvage the quartessence scenario.

A minimally coupled k-essence model \cite{arm15,garr47}, is described by
\begin{equation}
S = \int \, d^{4}x \, \sqrt{- g}  \left[ - \frac{R}{16\pi G} + {\cal L} (\varphi,X) \right],
\label{eq4001}
\end{equation}
where ${\cal{L}}$ is the most general Lagrangian, 
which depends on a single  scalar field
$\varphi$ of dimension $m^{-1}$, and on
the dimensionless quantity $X$ defined in (\ref{eq1203}).
For $X>0$  the energy momentum tensor
obtained
from 
(\ref{eq4001})
takes the perfect fluid
form, 
\begin{equation}
T_{\mu \nu}= 2{\cal L}_{X}\:
\varphi_{,\mu}\varphi_{,\nu}
-{\cal L} g_{\mu\nu}=(\rho+p) u_\mu u_\nu - p\, g_{\mu\nu}\, ,
\label{eq03}
\end{equation}
with ${\cal{L}}_{X}$ denoting $\partial {\cal{L}}/\partial X$ and 4-velocity
\begin{equation}
u_\mu = {\rm sgn}\, (\varphi_{,0})\frac{\varphi_{,\mu}}{\sqrt{X}}\, .
\label{eq511}
\end{equation}
The sign of $u_\mu$
  is chosen so $u_{0}$ is positive.
The associated hydrodynamic quantities  are
\begin{equation}
p ={\cal L}(\varphi,X);
\hspace{.5in}
\rho = 2 X {\cal L}_{X}(\varphi,X)-{\cal L}(\varphi,X),
\label{eq4004}
\end{equation}
and the speed of sound is defined as \cite{bilic}
\begin{equation}
c_s^{2} \equiv \left.\frac{\partial p}{\partial\rho}\right|_{s/n}= \left.\frac{\partial p}{\partial\rho}\right|_\varphi = \frac{{\cal{L}}_{X}}
{{\cal{L}}_{X} + 2 X {\cal{L}}_{XX}} .
\label{eq4006}
\end{equation}
Two general conditions 
${\cal{L}}_{X} \geq 0 $ 
and
${\cal{L}}_{XX} \geq 0 $ are
required for stability \cite{hsu48} and causality \cite{ell62}.
Now, using (\ref{eq511})-(\ref{eq4004}) the $\varphi$ field equation 
can be expressed as
\begin{equation}
\dot{\rho} + 3 {\cal{H}} ( \rho + p) + ( \dot{\varphi} -{\rm sgn}\,
(\varphi_{,0})\sqrt{X} )\partial {\cal{L}}/\partial\varphi =0.
\label{eq025}
\end{equation}

Since the 4-velocity 
(\ref{eq511}) 
is derived from a potential, the associated rotation tensor vanishes identically. The Raychaudhuri equation for the velocity congruence 
combined with Einstein's equations and the Euler equation assumes a simple form
\begin{equation}
3\dot{\cal H}+3{\cal H}^2
+\sigma_{\mu\nu}\sigma^{\mu\nu}
+4\pi G (\rho +3p)=
\left(\frac{c_s^2 h^{\mu\nu} \rho_{,\nu}}{p+\rho}\right)_{;\mu}\, ,
\label{eq008}
\end{equation}
 where $\sigma_{\mu\nu}$ is the shear tensor and
 $h_{\mu\nu}=g_{\mu\nu}-u_\mu u_\nu$
is a projector onto the three-space orthogonal to $u^\mu$.
The quantity ${\cal H}$ is the local Hubble parameter.
defined as
$ 3 {\cal H}  ={u^\nu}_{;\nu}$.
We thus obtain an evolution equation for ${\cal{H}}$ 
 sourced by
shear, density, pressure and pressure gradient. 
If $c_s = 0$, as for dust, Eq.~(\ref{eq008})
 and the continuity equation comprise the spherical model \cite{gazt35}.
However, we are not
interested in dust, since generally $c_s \neq 0$ and the right hand side of
(\ref{eq008}) is not necessarily zero.

 In general, the 4-velocity $u^{\mu}$ can be decomposed as \cite{bil49}
\begin{equation}
u^{\mu} = \left( U^{\mu} + v^{\mu} \right) / \sqrt{1 - v^{2}} \; \; ,
\label{eq031}
\end{equation}
where $U^{\mu} = \delta_{0}^{\mu} / \sqrt{g_{00}}$ is the 4-velocity  of fiducial observers at rest, and $v^{\mu}$ is spacelike, with $v^{\mu} v_{\mu} = - v^{2}$ and $U^{\mu} v_{\mu} = 0$.
In comoving coordinates  $v^\mu=0$.

In spherically symmetric spacetime it is convenient to write the metric in the form
\begin{equation}
ds^2 =N(t,r)^2dt^2 - b(t,r)^2(dr^2+r^2f(t,r)d\Omega^2), 
\label{eq09}
\end{equation}
where $N(t,r)$ is the lapse function, $b(t,r)$ is the local expansion scale,
and $f(t,r)$ describes the departure from the flat space for which $f=1$.
We assume that $N$, $a$, and $f$  are arbitrary functions of $t$ and $r$ which are
regular
and different from zero  at $r=0$.
Then, the local Hubble parameter and the shear are given by
\begin{equation}
{\cal H}=\frac{1}{N}\left(\frac{b_{,0}}{b}+
\frac{1}{3}\frac{f_{,0}}{f}\right);
\hspace{0.3in}
\sigma_{\mu\nu}\sigma^{\mu\nu}
=\frac{2}{3}\left(\frac{1}{2N} \frac{f_{,0}}{f}\right)^2 .
\label{eq160}
\end{equation}
In addition to the spherical symmetry we also require 
an FRW spatially flat asymptotic geometry, i.e., 
for $r\rightarrow \infty$ we demand 
 \begin{equation}
N\rightarrow  1; \hspace{.5cm} f\rightarrow 1;
\hspace{.5cm}
b\rightarrow {a}(t),
\label{eq091}
\end{equation}
where $a$ denotes the background expansion scale.

The righthand side of (\ref{eq008}) is difficult to treat in full generality.
As in \cite{bil34}, we apply the ``local approximation". The density contrast $\delta = (\rho - \bar{\rho})/\bar{\rho}$ is assumed to be of fixed Gaussian shape of comoving size $R$ with time-dependent amplitude, so that
\begin{equation}
    \rho (t,r) = \bar{\rho}(t)[1+ \delta_{R}(t)
  \,  e^{-r^2/(2 R^2)}],
\label{eq308}
\end{equation}
and the spatial derivatives are evaluated at the origin. This is in keeping with the spirit of the spherical model, where each region is treated as independent.
Since  $\partial_i \rho=0$ at $r=0$,
naturally $\partial_i N=0$ 
and $\partial_i b=0$ at $r=0$.
Hence,
\begin{equation}
 N(t,r)= N(t,0)(1+{\cal{O}} (r^2)); \hspace{1cm}
 b(t,r)=b(t,0)(1+{\cal{O}} (r^2)).
\end{equation}
Besides,  one finds $f_{,0}\rightarrow 0$ as  $r\rightarrow 0$ which  follows from
 Einstein's equation ${G^1}_0=0$.

From now on we denote by ${\cal H}$, $b$, and $N$ the corresponding functions
of $t$ and $r$ evaluated at $r=0$, i.e.,
 ${\cal H}\equiv {\cal H}(t,0)$,
$b\equiv b(t,0)$ and $N\equiv N(t,0)$.
 According to (\ref{eq160}), the shear scalar $\sigma_{\mu\nu}\sigma^{\mu\nu}$ vanishes at the origin. 
Evaluating (\ref{eq008}) at $r=0$
yields our working approximation to
the Raychaudhuri equation.

We will now apply our formalism to a particular subclass of k-essence unification 
models described by  (\ref{eq01}).
 The equation of state is then given by 
\begin{equation}
 p = 
- \frac{V(\varphi)^2}{\rho}\, ,
\label{eq049a}
\end{equation}
and the quantity $X$ may be expressed as
\begin{equation}
X(\rho, \varphi) = 1-\frac{V(\varphi)^2}{\rho^2} = 1 - c_{s}^{2} = 1 + w .
\label{eq050}
\end{equation}

The continuity equation, Eq.\ (\ref{eq025}), and Eq.\ (\ref{eq008}) evaluated at $r=0$
determine the
evolution of the density contrast. However, this set of equations is not complete as
it must be supplemented by a similar set of equations for the background
quantities  $\bar{\rho}$ and $H$.
The complete set of equations for $\bar{\rho}$, $H$, $\varphi$, 
$b$, $\rho$, and $\cal H$ is 
\begin{equation}
\left(\frac{d\varphi}{dt}\right)^2 = X (\varphi,\bar{\rho}),
\label{eq2}
\end{equation}
\begin{equation}
 \frac{d \bar{\rho}}{dt} + 3 H(\bar{\rho}+\bar{p})=0, 
\label{eq17}
\end{equation} 
\begin{equation}
\frac{d H}{dt}+H^2
+\frac{4\pi G}{3} (\bar{\rho} +3\bar{p})=0,
\label{eq18}
\end{equation}
\begin{equation}
\frac{db}{dt}=N b{\cal H},
\label{eq161}
\end{equation}
\begin{equation}
\frac{d\rho}{dt} + 3 N \; {\cal{H}} \; (\rho + p) = 0,
\label{eq4}
\end{equation}
\begin{equation}
\frac{d\cal{H}}{dt} + N \; \left[ {\cal{H}}^{2} +
\frac{4 \pi G}{3} \; (\rho + 3 p) -
\frac{c_{s}^{2} \; (\rho - \bar{\rho})}{b^{2} R^{2} (\rho + p)} \right] = 0,
\label{eq5}
\end{equation}
where $\bar{p}=p(\bar{\rho},\varphi)$ and $N=\sqrt{ X (\varphi, \bar{\rho}) / X (\varphi, \rho )} $. 
Eqs.\ (\ref{eq2}) and (\ref{eq161}) follow from (\ref{eq025}) and (\ref{eq160}),
respectively,
 Eqs.\ (\ref{eq025}) and (\ref{eq4}) are
the continuity equations, and Eqs.\ (\ref{eq18}) and (\ref{eq5}) are the Raychaudhuri equations for the background and the spherical inhomogeneity, respectively.

Now we restrict our attention to the potential (\ref{eq08}). In the high density regime we have $X \simeq 1$, and (\ref{eq2}) can be integrated yielding 
$\varphi \simeq 2/(3 H)$. Here $H \simeq H_{0} \sqrt{\Omega} a^{-3/2}$ with $\Omega$ being the equivalent matter content at high redshift. Hence, $V(\varphi)^2 \sim a^{6n}$, which leads to a  suppression 
 of 10$^{-6}$ of the acoustic horizon
at $z = 9$ for $n = 1$.

 To proceed we require a value for the constant $V_{n}$ in the potential (\ref{eq08}). 
 As the main purpose of this paper is to investigate the evolution of inhomogeneities
we will not pursue the exact fitting of the background evolution.
 Instead,
we estimate  $V_{n}$ as follows. We integrate (\ref{eq2}) approximately with 
\begin{equation}
X=1+w(a)\simeq 1-\frac{\Omega_\Lambda}{\Omega_\Lambda+\Omega a^{-3}}\, ,
\hspace{1cm}
 \Omega + \Omega_{\Lambda} = 1,
\end{equation}
as in a $\Lambda$CDM universe \cite{bert44} and 
we fix the pressure given by (\ref{eq01}) to equal that of $\Lambda$ at $a = 1$.
In this way
the naive background in our model reproduces the standard cosmology 
from decoupling up to the scales of about $a=0.8$ and 
fits the cosmology today only approximately (figure \ref{fig1}(a)).

We solve our differential equations with
$a$ starting from the initial $a_{\rm dec}=1/(z_{\rm dec}+1)$  
at decoupling  redshift $z_{\rm dec}=1089$ for a particular
comoving size $R$. 
The initial values for the background are given by
\begin{equation}
\bar{\rho}_{\rm in}=\rho_0\frac{\Omega}{a_{\rm dec}^3};
 \hspace{1cm}
H_{\rm in}=H_0\sqrt{\frac{\Omega}{a_{\rm dec}^3}};
\hspace{1cm}
\varphi_{\rm in} = \frac{2}{3 H_{\rm in}},
\label{eq23}
\end{equation}
and for the initial inhomogeneity we take
\begin{equation}
\rho_{\rm in}=\bar{\rho}_{\rm in} (1+\delta_{\rm in}) \; , 
 \hspace{0.75cm}
{\cal H}_{\rm in}=H_{\rm in}\left(1-\frac{\delta_{\rm in}}{3}\right) \; ,
\end{equation}
where $\Omega=0.27$ represents the effective dark matter fraction
and $\delta_{\rm in}=\delta_R (a_{\rm dec})$ is a variable initial density
contrast, chosen arbitrarily for a particular $R$.

\def\figsubcap#1{\par\noindent\centering\footnotesize(#1)}
\begin{figure}[t]%
\begin{center}
  \parbox{2.1in}{\epsfig{figure=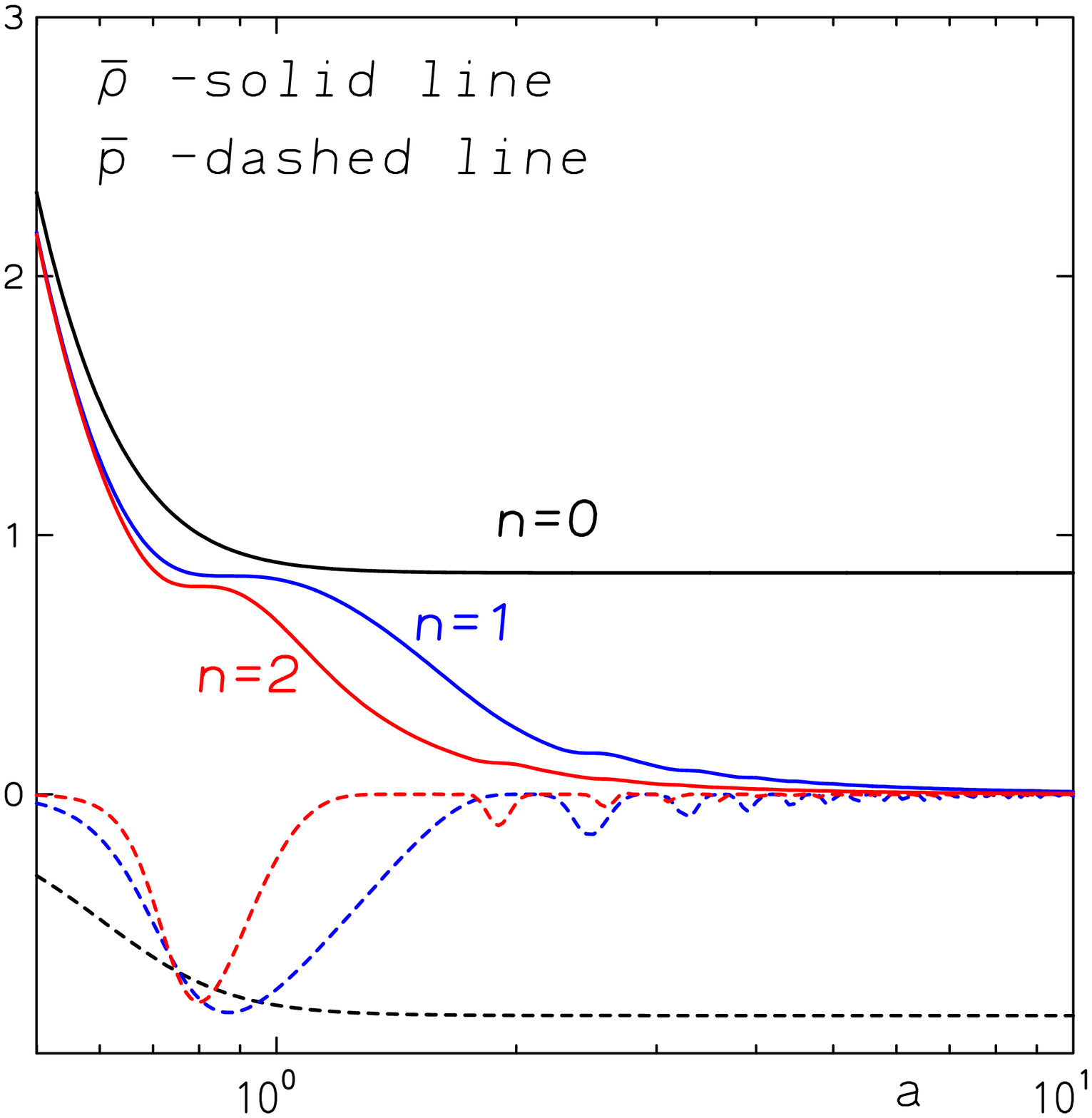,width=2in}\figsubcap{a}}
  \hspace*{4pt}
  \parbox{2.1in}{\epsfig{figure=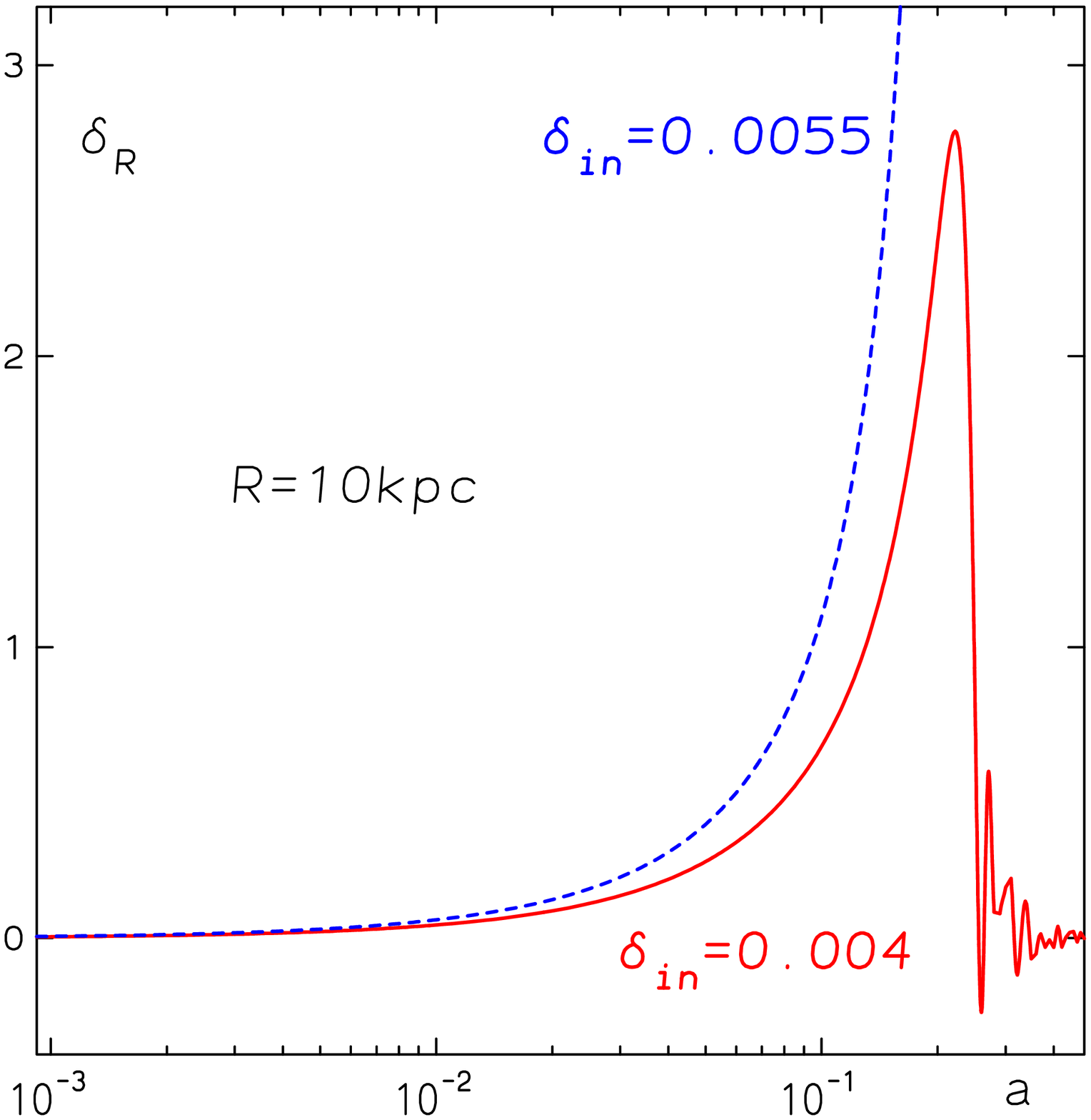,width=2in}\figsubcap{b}}
\caption{
(a) Evolution of the background in
            the tachyon spherical model.
(b) Evolution of $\delta_{R}(a)$ 
           from $a_{\rm dec} = 1/1090$
           for $R$ = 10 kpc,
           $\delta_{R} (a_{\rm dec})$ =0.004 (solid)
            and $\delta_{R} (a_{\rm dec})$ =0.0055 (dashed).
 }
\label{fig1}
\end{center}
\end{figure}


In figure \ref{fig1}(b) the representative case of evolution of two initial
perturbations starting from 
decoupling
for $R$ = 10 kpc is shown  for $n = 2$.
 The plots  represent two distinct regimes: 
the growing mode or condensation (dashed line) and the damped oscillations
( solid line).
In contrast to the linear theory, where for any $R$ the acoustic
horizon will eventually stop $\delta_R$ from growing, irrespective of the initial
value of the perturbation, here we have for an
initial $\delta_{R} (a_{\rm dec})$ above a certain
threshold $\delta_{c} (R)$,
$\delta_{R} (a) \rightarrow \infty$ at
finite $a$, just as in the dust model.
 Thus perturbations with $\delta_{R} (a_{dec}) \geq \delta_{c} (R)$ evolve into
a {\it nonlinear} gravitational condensate that at low $z$ behaves as pressureless super-particles.
Conversely, for a sufficiently small
$\delta_{R} (a_{\rm dec})$, the acoustic horizon can stop
$\delta_{R} (a)$ from growing.

The crucial question now  is what fraction of the tachyon gas goes into condensate.
In \cite{bil31} it was shown that if this fraction was sufficiently large, the
CMB and the mass power spectrum
could be reproduced for the simple Chaplygin gas.
To answer this question quantitatively, 
we follow the  Press-Schechter
procedure \cite{pres53} as in  \cite{bil34}.
Assuming $\delta_{R}(a_{\rm dec})$ is given by a Gaussian random field with
dispersion $\sigma(R)$, 
the condensate fraction at
a scale $R$ is given by
\begin{equation}
F (R) = 2 \int_{\delta_{c}(R)}^{\infty} \; \frac{d \delta}{ \sqrt{2 \pi} \sigma (R) }\;
{\rm exp} \left( - \frac{\delta^{2}}{2 \sigma^{2}(R)} \right)
= {\rm erfc} \left(
\frac{ \delta_{c} (R) }{ \sqrt{2} \; \sigma (R) }
\right) \, ,
\label{eq401}
\end{equation}
where $\delta_c(R)$ is the threshold 
shown in figure \ref{fig2}(a) . In figure \ref{fig2}(a) we also exhibit
the dispersion
\begin{equation}
\sigma^{2}(R) = \int_{0}^{\infty} \; \frac{dk}{k} \; {\rm exp}
( - k^{2} R^{2} ) \Delta^{2} (k, a_{\rm dec}) ,
\label{eq402}
\end{equation}
calculated using  the 
variance of the concordance
model
\cite{hin3} 
\begin{equation}
\Delta^{2} (k,a)={\rm const} \left(\frac{k}{a H}\right)^4
T^2(k)\left(\frac{k}{7.5 a_0 H_0}\right)^{n_{s}-1}\, .
\label{eq403}
\end{equation}
In figure \ref{fig2}(b) we present $F(R)$ for
const=7.11$\times 10^{-9}$, the spectral index $n_{s}$=1.02,
and the parameterization of Bardeen et al.\ \cite{bard54} for the transfer
function $T(k)$ with $\Omega_{B}$=0.04.
The parameters are fixed by fitting (\ref{eq403})
to the 2dFGRS power spectrum data \cite{perc55}.
Our result
demonstrates that the collapse fraction is about
70\% for $n=2$ for a wide range of the comoving size $R$
 and peaks at about 45\% for $n = 1$.

\def\figsubcap#1{\par\noindent\centering\footnotesize(#1)}
\begin{figure}[t]%
\begin{center}
  \parbox{2.1in}{\epsfig{figure=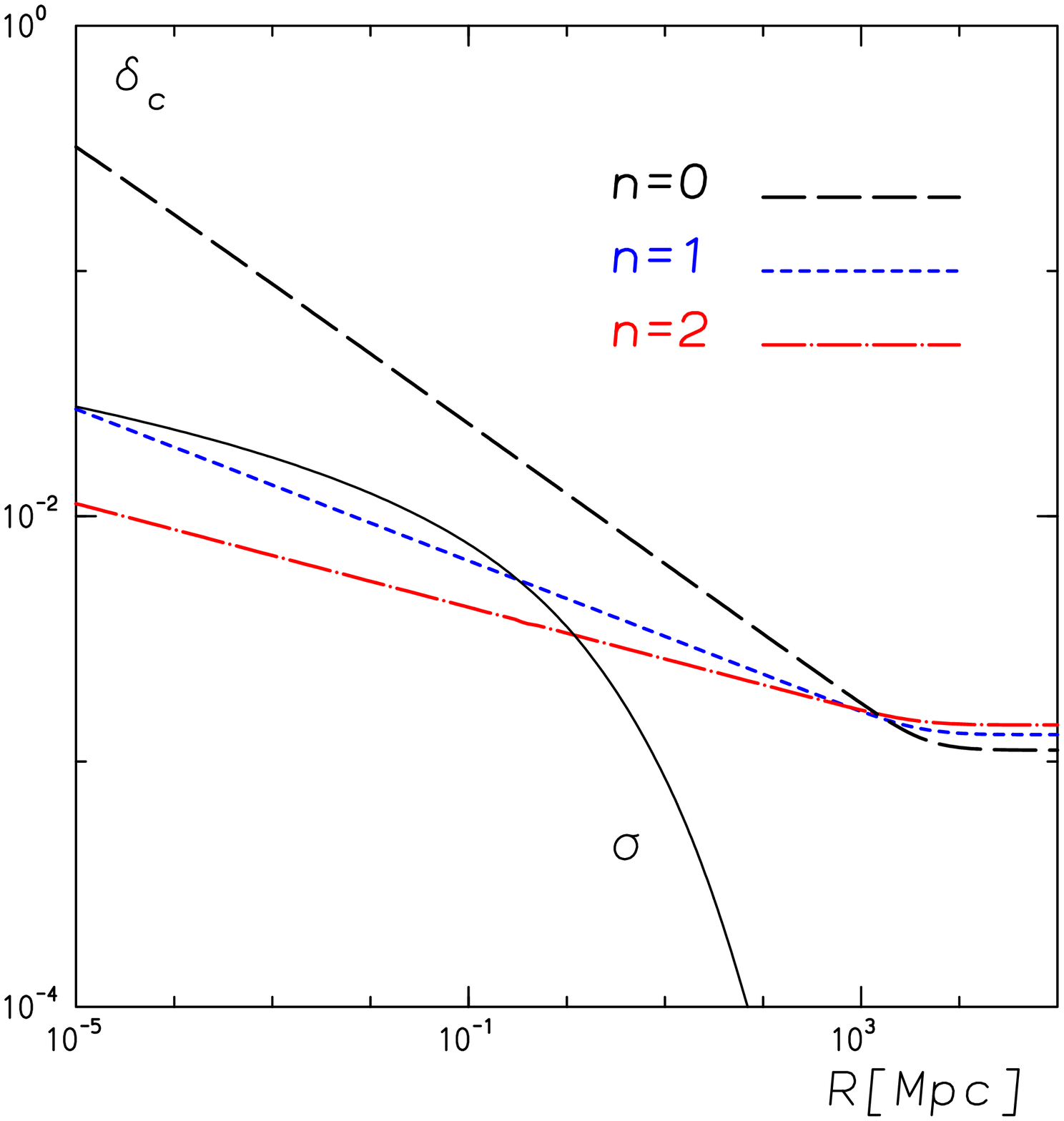,width=2in}\figsubcap{a}}
  \hspace*{4pt}
  \parbox{2.1in}{\epsfig{figure=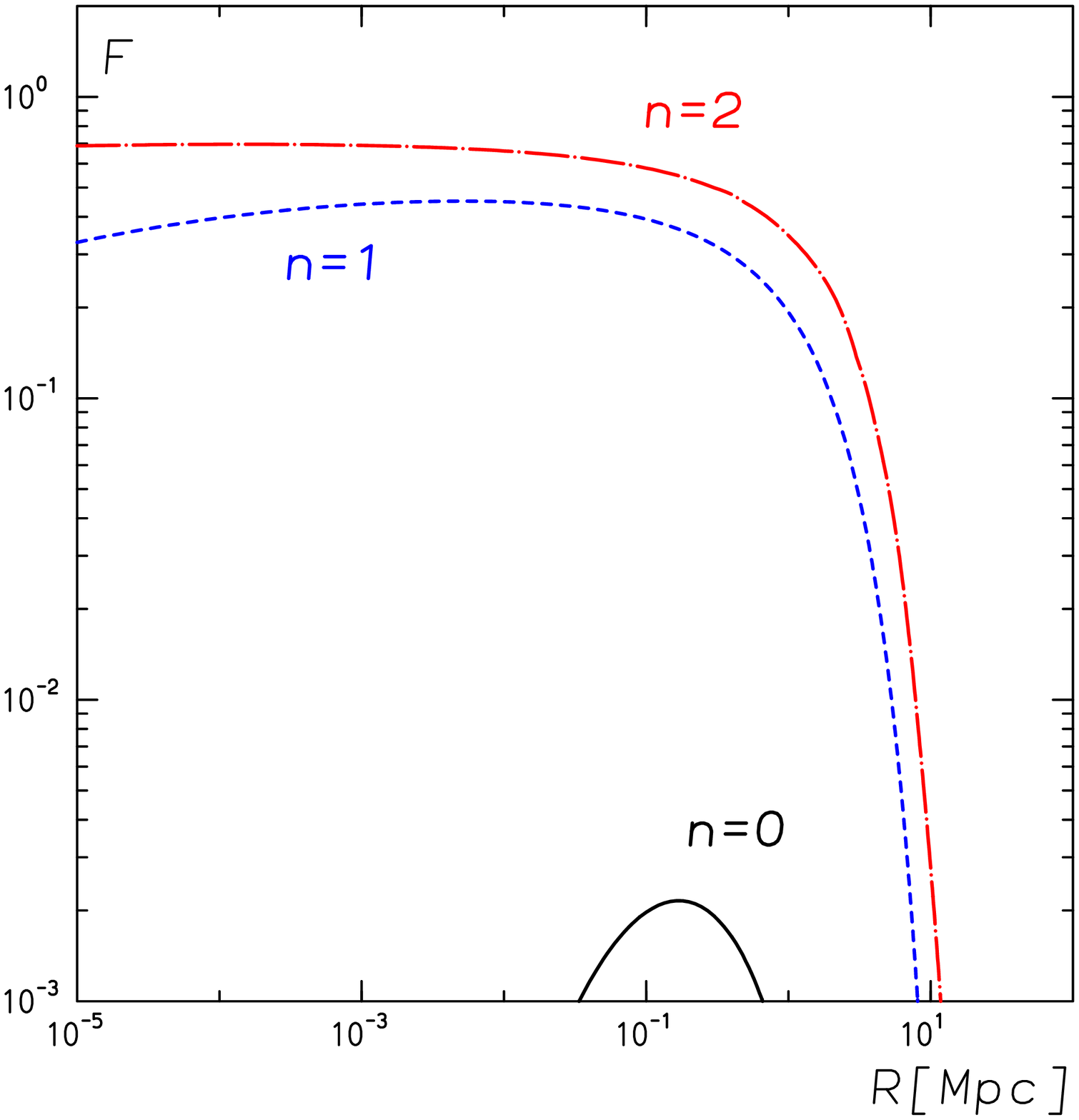,width=2in}\figsubcap{b}}
\caption{
(a) Initial value $\delta_{R}(a_{\rm dec})$  versus $R$
for $\Omega=0.27$ and $h=0.71$.
The threshold $\delta_c (R)$
is shown by the line separating
 the condensation regime from the damped oscillations regime.
The solid line gives $\sigma (R)$ calculated using the
concordance model.
(b) Fraction of the tachyon gas in collapsed objects using
$\delta_c(R)$ and $\sigma(R)$ depicted in
 (a).}
  \label{fig2}
\end{center}
\end{figure}

Albeit encouraging, these preliminary results do not in themselves demonstrate that 
the tachyon with potential (8) constitutes a  viable cosmology. Such a step requires the inclusion of baryons and comparison with the full cosmological data. What has been shown is that it is not correct in an adiabatic model to simply pursue linear perturbations to the original background: the system evolves nonlinearly into a mixed system of gravitational condensate and residual k-essence so that the ``background'' at low $z$ is quite different from the initial one. 
Because of this one needs new computational tools for a meaningful confrontation with the data. 
%


The  tachyon k-essence unification remains to be tested against large-scale structure and CMB observations.  An encouraging feature of the positive power-law potential is that it provides for acceleration as a periodic transient phenomenon 
\cite{fro58}
which obviates the de Sitter horizon problem \cite{bil37}.

\section*{Acknowledgments}
We wish to thank Robert Lindebaum for useful discussions.
This research is in part supported by
the Foundation for Fundamental Research
(FFR) grant number  PHY99-1241, the National Research Foundation of South Africa grant number FA2005033 100013, and the Research Committee of the
University of Cape Town.
The work of NB is supported
in part by the Ministry of Science and Technology of the
Republic of Croatia under Contract No. 098-0982930-2864.
 


\begin{thebibliography}{99}
%

%
\bibitem{kam9} A. Kamenshchik, U. Moschella, and V. Pasquier, {\em Phys. Lett. B} {\bf 511}, 265 (2001).
\bibitem{bil7} N. Bili\'c, G.B. Tupper, and R.D. Viollier,
{\em Phys. Lett. B} {\bf 535}, 17 (2002).
\bibitem{bil8} N. Bili\'c, G.B. Tupper, and R.D. Viollier,
in {\it  DARK 2002},
eds.  H.V. Klapdor-Kleingrothaus and R.D. Viollier, {\em Int.\ Conf.\ B} {\bf 535}, 17 (2002);
[arXiv:astro-ph/0207423].
\bibitem{jack10} R.~Jackiw,
{\it Lectures on fluid dynamics}
(Springer-Verlag, New-York, 2002).
\bibitem{mak11} M. Makler, S.Q. de Oliveira, and I. Waga,
{\em Phys. Lett. B} {\bf 555}, 1 (2003).
\bibitem{mak12} R.R.R. Reis, M. Makler, and I. Waga,
{\em Phys. Rev. D} {\bf 69}, 101301 (2004).
\bibitem{chi13} L.P. Chimento,  {\em Phys. Rev. D} {\bf 69}, 123517 (2004).
\bibitem{sch14} R.J. Scherrer, {\em Phys. Rev. Lett.} {\bf 93}, 011301 (2004).
\bibitem{arm15} C. Armendariz-Picon, T. Damour, and V. Mukhanov,
{\em Phys. Lett. B} {\bf 458}, 209 (1999).
\bibitem{fab19} J.C. Fabris,  S.V.B. Gon\c calves, and P.E. de Souza,
{\em Gen. Relativ. Gravit.} {\bf 34}, 53 (2002); ibid.\
 {\bf 34}, 2111 (2002).
\bibitem{gor20} V. Gorini, A.Y. Kamenshchik, U. Moschella, O.F. Piattella, and
A.A. Starobinsky,  {\em JCAP} {\bf 0802}, 016 (2008).
\bibitem{sand21} H.B. Sandvik, M. Tegmark,
M. Zaldarriaga, and I. Waga,
{\em Phys. Rev. D} {\bf 69}, 123524 (2004).
\bibitem{cart22} P. Carturan and F. Finelli,{\em Phys. Rev. D} {\bf 68}, 103501 (2003).
\bibitem{sen40} M.R. Garousi,
  {\em Nucl.\ Phys.\  B} {\bf 584}, 284 (2000);
A. Sen,
{\em  JHEP} {\bf 0207}, 065 (2002).
\bibitem{bil39}
N. Bili\'c, G.B. Tupper, and R.D. Viollier, {\em J. Phys. A} {\bf 40}, 6877 (2007);
[arXiv: gr-qc/0610104].
\bibitem{bilic}
 N. Bilic, G.B. Tupper and R.D. Viollier,
{\em Phys.\ Rev.\  D} {\bf 80}, 023515 (2009)
  [arXiv:0809.0375 [gr-qc]].
\bibitem{bil34} N. Bili\'c, R.J. Lindebaum, G.B. Tupper, and R.D. Viollier,
{\em JCAP} {\bf 0411}, 008 (2003); [arXiv: astro-ph/0307214].
\bibitem{guo43} Z.-K. Guo and Y.-Z. Zhang, {\em Phys. Lett. B} {\bf 645}, 326 (2007).
\bibitem{garr47} J. Garriga and V.F. Makhanov, 
{\em Phys. Lett. B} {\bf 458}, 219 (1999).
\bibitem{hsu48} S.D.H. Hsu, A. Jenkins, and M.B. Wise, 
{\em Phys. Lett. B} {\bf 597}, 270 (2004);
N. Bilic, G.B. Tupper, and R.D. Viollier, 
{\em JCAP} {\bf 0809}, 002 (2008); [arXiv: 0801.3942].
\bibitem{ell62} G.F.R. Ellis, R. Maartens, and M.A.H. MacCallum, {\em Gen. Relativ. Gravit.} {\bf 39}, 1651 (2007).
\bibitem{gazt35} E. Gazt\~{a}naga and J.A. Lobo, {\em Astrophys. J.} {\bf 548}, 47 (2001).
\bibitem{bil49} N. Bili\'c, {\em Class. Quant. Grav.} {\bf 16}, 3953 (1999); [arXiv: gr-qc/9908002].
\bibitem{bert44} D. Bertacca, S. Matarrese, and M. Pietroni, 
{\em Mod. Phys. Lett.} {\bf A22}, 2893 (2007).
\bibitem{bil31}
N. Bili\'{c}, R.J. Lindebaum, G.B. Tupper, and R.D. Viollier, in
{\it Proceedings of the XVth Rencontres de Blois, France, 2003},
eds.\ J. Dumarchez et al. (The Gioi Publishers, Vietnam, 2005);
[arXiv: astro-ph/0310181].
\bibitem{pres53} W.H. Press and P. Schechter, 
{\em Astrophys. J.} {\bf 187}, 425 (1974).
\bibitem{hin3} G. Hinshaw et al.,
{\em Astrophys. J. Suppl.} {\bf 170}, 288 (2007);
D.N. Spergel et al. {\em Astrophys. J. Suppl.} {\bf 170}, 377 (2007);
E. Komatsu et al., {\em Astrophys. J. Suppl} {\bf 180}, 330 (2009).
\bibitem{bard54} J.M. Bardeen, J.R. Bond, N. Kaiser, and A.S. Szalay, 
{\em Astrophys. J.} {\bf 304}, 15 (1986).
\bibitem{perc55} W.J. Percival et al., 
{\em Mon. Not. R. Astron. Soc.} {\bf 327}, 1297 (2001).
%
\bibitem{fro58} A. Frolov, L. Kofman, and A. Starobinsky,
{\em Phys.\ Lett.\ B} {\bf 545}, 8 (2002).
\bibitem{bil37}
N. Bili\'c, G.B. Tupper, and R.D. Viollier, 
{\em JCAP} {\bf 0510}, 003 (2005);
[arXiv: astro-ph/0503428].  

\end{thebibliography}
\end{document}